\documentstyle[epsf,12pt]{mrsproc}
\begin{document}
\begin{center}
{\bf Enhanced polarization in strained BaTiO$_3$ from first principles}
\end{center}

\vspace{0.2cm}
\noindent
{J. B. Neaton, C.-L. Hsueh, and K. M. Rabe} \\ 
{Department of Physics and Astronomy, Rutgers University\\
Piscataway, NJ 08854-8019}
\begin{abstract}
The structure, polarization, and zone-center phonons of 
bulk tetragonal BaTiO$_3$ under compressive epitaxial stress are 
calculated using density functional theory within the local density approximation. 
The polarization, computed using the Berry-phase formalism, increases 
with increasing tetragonality and is found to be enhanced by
nearly 70\% for the largest compressive misfit strain considered (-2.28\%). 
The results are expected to be useful for the analysis 
of coherent epitaxial BaTiO$_3$ thin films and heterostructures 
grown on perovskite substrates having a smaller lattice constant, such
as SrTiO$_3$.
\end{abstract}

\section{INTRODUCTION}

With recent advances in oxide epitaxy, coherent growth of 
nearly perfect ultrathin perovskite 
films and heterostructures is now possible \cite{schlom1,schlom2,ahn}. 
In these systems, individual constituent layers
are under mixed mechanical boundary conditions: constrained to the
substrate lattice constant in-plane, the films are free to relax in the
normal direction. The strains achieved in this way can be quite significant, and
can be controlled to tune structure, ferroelectric, and dielectric
properties of these nanostructured materials \cite{sup1,sup2}. 
For example, the polarization could be increased over that 
of known bulk compounds \cite{schimizu,pertsev1}, improving the performance of ferroelectric 
field-effect transistors and nonvolatile memories.

When the interfaces in lattice-matched heterostructures are between 
sufficiently similar materials, even the thinnest layers can be well described by considering
the behavior of the bulk under the same boundary conditions. 
Thus, for example, many properties of 
BaTiO$_3$/SrTiO$_3$ superlattices can be understood by
considering the effects of large strains
on bulk BaTiO$_3$. This has led to great interest in information about highly strained
bulk and film states in the perovskites, 
for example as presented previously in a phenomenological
Landau-Devonshire framework \cite{pertsev1,pertsev2,pertsev3}. 

Here we perform first-principles calculations and obtain information
about the structure, spontaneous polarization, and zone-center phonons
of bulk tetragonal BaTiO$_3$ with misfit strains down to about -2.3\%.
While our results correspond to zero temperature, previous work \cite{pertsev1} suggests
that in this high-strain regime,
the tetragonal phase is stable with increasing temperature and thus
our analysis is relevant to the study of ultrathin BaTiO$_3$
films and BaTiO$_3$/SrTiO$_3$ superlattices at room temperature \cite{us}.

\section{METHODOLOGY}

To predict the ground state structure of strained BaTiO$_3$ and compute the associated 
polarization, we use density functional theory (DFT) \cite{hk} within the local density
approximation (LDA) \cite{ks}, as implemented with a plane-wave basis in the Vienna {\it ab initio} 
Simulations Package (VASP) \cite{kresse}. Results are obtained using 
projector-augmented wave (PAW) \cite{paw} potentials provided with VASP. 
The PAW potentials explicitly treat 10 valence electrons 
for Sr ($4s^24p^65s^2$), 12 for Ti ($3s^23p^63d^24s^2$), and 6 for oxygen ($2s^22p^6$).
The ions are steadily relaxed toward equilibrium
until the Hellmann-Feynman forces are less than 10$^{-3}$ eV/{\AA}.
Brillouin zone integrations are performed with a
Gaussian broadening of 0.1 eV during all relaxations.  
All calculations are performed with a 6$\times$6$\times$6
Monkhorst-Pack {\bf k}-point mesh and a 50 Ry plane-wave cutoff. 
Once ground-state structures are obtained, polarizations are determined from
first principles using the Berry-phase formalism  \cite{ksv}.
The force-constant matrices are then obtained with a frozen-phonon approach.
Specifically, a series of calculations is performed in which each ion in turn is displaced along
$\hat z$ by 0.1\% of the lattice parameter and the Hellmann-Feynman forces on all
the atoms computed; the phonon frequencies and eigenvectors are then acquired through 
diagonalization of the corresponding dynamical matrix.

\section{RESULTS}

We begin by calculating the equilibrium structural parameters of
BaTiO$_3$ under zero stress. The observed phases of bulk BaTiO$_3$ are
closely related to the cubic perovskite structure $Pm\bar{3}m$  \cite{lg},
which is the structure of the high-temperature phase (T$>$403 K).
The simple cubic unit cell of this structure can be chosen so the Ti cation
occupies the body-centered site 1b ($1 \over 2$,$1 \over 2$,$1 \over 2$)
and is octahedrally coordinated by oxygen anions at 3c ($1 \over 2$,$0$,$1 \over 2$); 
the Ba cation is then at Wyckoff position 1a ($0$,$0$,$0$), and twelve-fold 
coordinated by oxygens.

The rhombohedral $P3m1$ phase is computed to be the ground state of BaTiO$_3$,
consistent with observations \cite{lg}; however, 
it is just slightly lower in energy (only $\sim$1 meV per unit cell)
than the optimized tetragonal $P4mm$ structure  \cite{jorge}, 
the observed room temperature phase \cite{lg}.
In the tetragonal phase, the simple cubic lattice distorts, resulting in two
independent lattice parameters $a$ (in-plane) and $c$ (normal). The lower symmetry
also permits relative displacement of the atomic sublattices along $\hat z$, with
Wyckoff positions 1a ($0$,$0$,$0$) for Ba and
1b ($1 \over 2$,$1 \over 2$,$1 \over 2$+$\Delta_{\rm Ti}$) for Ti,
while two Wyckoff positions are occupied by oxygens, designated
O$_{\rm I}$, and O$_{\rm II}$:
1b ($1 \over 2$,$1 \over 2$,$\Delta_{\rm O_{\rm I}}$)
and 2c ($1 \over 2$,$0$,$1 \over 2$+$\Delta_{\rm O_{\rm II}}$).
These displacements result in a nonzero polarization along [001].

The computed values of the structural parameters are given in the first line of Table I.
Our calculated lattice parameters
$a$=3.945~{\AA} and $c$=3.988~{\AA} are slightly 
smaller than those found experimentally ($a$=3.994~{\AA} 
and $c$=4.036~{\AA}  \cite{shirane}),
the size of the underestimate being typical of that found in LDA
studies of a wide variety of insulators and semiconductors.
The smaller volume is known, however, to suppress the
polar instability \cite{volume}, and indeed the 
computed displacements (in reduced units of $c$) 
of 0.0122, -0.0192, and -0.0124 for $\Delta_{\rm Ti}$,  
$\Delta_{\rm O_{\rm I}}$, and $\Delta_{\rm O_{\rm II}}$, respectively, 
are smaller than the measured values \cite{shirane}, 
which are 0.015, -~0.023, and -0.014.
The polarization of the optimized tetragonal phase ($P_0$=24.97~$\mu$C/cm$^2$)
is thus also significantly less than that computed with the
experimental structural parameters (32.2~$\mu$C/cm$^2$).
As a consequence, in the following we define misfit strains to be 
relative to the calculated lattice constants, and define
polarization enhancement relative to the computed zero-stress polarization $P_0$.

In coherent epitaxial thin films and [001] superlattice heterostructures,
the two lattice vectors in the plane of the interface are constrained to 
match the substrate while the third lattice vector is free to relax.
Here, we consider tetragonal BaTiO$_3$ ($P4mm$) under 
mechanical boundary conditions corresponding to coherent epitaxy of a $c$-axis
oriented phase on simple
cubic substrates, such as SrTiO$_3$, having lattice constants slightly
smaller than that of stress-free bulk tetragonal BaTiO$_3$.
We perform a series of constrained relaxations in which 
the in-plane lattice constant $a$ of bulk tetragonal BaTiO$_3$ is 
compressed (e.g., to the value of a hypothetical substrate) and the length of 
the normal axis and atomic positions are optimized. We consider misfit strains 
ranging from -0.28\% to -2.28\%.
The absolute stabilities of the resulting structures require independent investigation.
For example, it may be energetically favorable for the polarization vector and c-axis to
tilt toward the [110] direction
(this has been referred to as the $r$ phase in previous work  \cite{pertsev1}),
lowering the point symmetry to monoclinic. 
We have, however, verified that for misfit strains of -0.76\% and larger, lowering the symmetry to monoclinic
does not result in a lower energy, and we can therefore tentatively conclude that the $P4mm$ structure
is relevant under the boundary conditions considered in this work.

The optimized structural parameters of strained $c$-phase BaTiO$_3$ are summarized 
in Table I. A tendency to conserve volume is manifested as an
increase in $c$ with increasing magnitude of the misfit strain, through this increase
does not result in a full compensation of the in-plane compression and the volume actually decreases moderately.
The atomic displacements relative to the centrosymmetrically-strained cubic perovskite structure exhibit 
a significant, nearly linear, increase with
increasing misfit strain. Both the average and deviation (or splitting) of the two Ti-O$_{\rm I}$ distances
(along the $\hat z$-oriented Ti-O chain) increase with misfit strain, from
1.994 $\pm$ 0.125 {\AA} at zero misfit to 2.066 $\pm$ 0.233 {\AA} at -2.28\%. Thus,
the smaller Ti-O$_{\rm I}$ distance actually decreases slightly, from 1.87 to 1.83 {\AA},
while the larger distance increases by 8\% over the same range.
The shortest Ba-O$_{\rm II}$ distance also declines modestly from 2.77 to 2.75 {\AA}. 
The shortest Ba-O$_{\rm I}$ distance, however, 
decreases more rapidly with increasing strain and, for the largest 
strain considered (2.28\%), this separation is smaller than the minimum
Ba-O$_{\rm II}$ contact.

These structural trends are accompanied by a significant change in polarization $P_s$.
Specifically, the increased atomic displacements associated with the increase of tetragonality ($c/a$)
tend to increase the polarization; for 2.28\% strain, the polarization
{\it increases} by nearly a factor of two. These results are consistent with a previous first-principles
study, in which a strong dependence of the polarization on the $c$ parameter 
was observed for BaTiO$_3$~\cite{schimizu}. We also report the increase in energy per unit cell,
$\Delta E$, which reflects the energy cost of nonzero misfit strain. It varies
quadratically with misfit strain over the range shown, with an estimated
effective elastic constant of 135 eV/unit cell.

The computed frequencies for the $\hat z$ polarized zone-center phonons for each
optimized structure are given in Table II. As there are five atoms in each primitive cell,
12 zone-center optical phonons are expected (plus three acoustic modes).
At the zone center these are labeled by the irreducible representations of the C$_{4v}$ point group.
There are 4 A$_1$ modes (Raman and IR-active),
1 B$_1$ mode (Raman active), and 5 doubly-degenerate E modes (Raman and IR-active),
of which the A$_1$ and B$_1$ modes have atomic displacements and polarization
(if nonzero) only along $\hat z$.
The frequencies of the three A$_1$
modes are 503, 252, and 178~cm$^{-1}$, respectively. These agree quite well with
a room-temperature experiment, which reports 512, 276, and 178~cm$^{-1}$ for the 
same TO modes \cite{porto}. An analysis of the corresponding eigenvectors 
indicates that the two highest-frequency modes involves primarily Ti and O motion; 
in the lowest the cations move opposite to the anions, with each ion in the primitive cell 
participating almost equally. 
The B$_1$ mode, for which the computed frequency is 294~cm$^{-1}$, is Raman active and involves 
alternating motion of the two O$_{\rm II}$ ions. As for the A$_1$ modes, our
computed frequency agrees well with the value of 308~cm$^{-1}$ 
measured at room temperature \cite{porto}. 
We are presently unaware of previous first-principles 
calculations of zone-center phonon frequencies in tetragonal BaTiO$_3$, 
although we note previous calculations have been performed using an empirical
rigid-ion model \cite{freire} and shell-model potentials \cite{khalal}.

In the remainder of Table II, we report the effect of 
in-plane compression on the frequencies of these four modes.
We find that the two highest A$_1$ modes {\it increase} with 
increasing misfit strain, while the frequency of the lowest-lying A$_1$ mode
is nearly independent of misfit strain down to about -2.3\%. 
In contrast, the computed Raman-active B$_1$ mode frequency is found to 
{\it decrease} with increasing misfit strain. 
\begin{table}
\begin{center}
\caption{\small Structural parameters of BaTiO$_3$ under in-plane compression,
computed from first principles. 
The misfit strain (\%) is measured relative to the calculated 
lattice parameter of the computed tetragonal phase, reported in the table as 0.0\% strain. Displacements
are given relative to the centrosymmetrically-strained cubic-perovskite structure with the Ba
position fixed at (0,0,0) 
in reduced coordinates (units of $c$). P$_s$ is the 
Berry-phase polarization in $\mu$C/cm$^2$, with P$_0$ being the polarization of the equilibrium tetragonal structure. $\Delta E$ is the increase in
energy in meV per unit cell relative to the equilibrium tetragonal structure.}
\vspace{0.15cm}
\begin{tabular}{c c c c c c c c c c c}
\% & $a$ (\AA) & $c$ (\AA) & $c/a$ & V (\AA$^3$)& $\Delta_{\rm Ti}$ 
& $\Delta_{\rm O_{\rm I}}$ & $\Delta_{\rm O_{\rm II}}$ & P$_s$  
& P$_s$/P$_0$ & $\Delta$E  \\\hline
0.0 & 3.945 & 3.988  & 1.0110 & 62.06 & 0.0122 & -0.0192 & -0.0124 & 24.97 & 1    & 0.0\\
-0.28 & 3.934 & 4.0097 & 1.0192 & 62.06 & 0.0134 & -0.0227 & -0.0143 & 28.08 & 1.13 & 1.08\\
-0.76 & 3.915 & 4.0353 & 1.0309 & 61.83 & 0.0148 & -0.0256 & -0.0165 & 31.36 & 1.26 & 8.48\\
-1.26 & 3.895 & 4.0647 & 1.0437 & 61.66 & 0.0160 & -0.0295 & -0.0191 & 34.87 & 1.40 & 22.67\\
-1.77 & 3.875 & 4.0961 & 1.0571 & 61.51 & 0.0170 & -0.0336 & -0.0220 & 38.37 & 1.54 & 43.63\\
-2.28 & 3.855 & 4.1322 & 1.0719 & 61.42 & 0.0179 & -0.0384 & -0.0253 & 42.12 & 1.69 & 71.26\\
\end{tabular}
\end{center}
\end{table}

\begin{table}
\begin{center}
\caption{\small Computed zone-center phonon frequencies 
of BaTiO$_3$ as a function of misfit strain (\%) for $\hat z$-polarized phonons 
labelled by irreducible representations A$_1$ and B$_1$.}
\vspace{0.15cm}
\begin{tabular}{c c c c c}
\multicolumn{1}{c}{\%} &
\multicolumn{3}{c}{$\omega_{\rm A_1}$ (cm$^{-1}$)} &
\multicolumn{1}{c}{$\omega_{\rm B_1}$ (cm$^{-1}$)}\\
\hline
0.0 & 503 & 252 & 178 & 294 \\
-0.28 & 508 & 275 & 179 & 292 \\
-0.76 & 514 & 296 & 180 & 290 \\
-1.26 & 524 & 316 & 180 & 286 \\
-1.77 & 536 & 332 & 180 & 282 \\
-2.28 & 549 & 347 & 180 & 278 \\
\end{tabular}
\end{center}
\end{table}

\section{DISCUSSION}

The trends witnessed here in the computed properties of strained BaTiO$_3$
are quite general and reflect the underlying physics of ferroelectric perovskites under 
the mixed-mechanical boundary conditions associated with epitaxy.
In-plane compression resulting from epitaxial stress favors expansion 
of $c$ normal to the substrate and, by a mechanism analogous
to that responsible for the enhancement of the ferroelectric instability with increasing volume, 
the polar distortion and associated polarization along $\hat z$ is expected to increase,
as observed.

While larger polarization enhancements may in principle be achieved through 
larger misfit strains, the role of misfit strain in materials design is 
limited by a rapidly increasing 
strain energy $\Delta E$. Above a critical thickness, misfit dislocations 
will become energetically favorable, and the film or heterostructure will
relax to its bulk lattice constant, relieving stress and decreasing the
polarization. 
However, coherent epitaxy of BaTiO$_3$-based heterostructures of useful thicknesses
has already been demonstrated \cite{schlom2,sup2}, and the present results show that 
the misfit strain of BaTiO$_3$ on a SrTiO$_3$
substrate ($\sim$2\%) is large enough to achieve significant enhancement of the polarization.

Our computed phonon frequencies are of interest chiefly for two reasons. 
First, combined with the computed Born effective charges and electronic dielectric constant
(to be reported in a separate publication), they
determine the phonon contribution to the static dielectric response $\epsilon_{0}$. While clearly an
important property in itself, $\epsilon_0$ should also be relevant for understanding
the effects of changing electrical boundary conditions on films and superlattices.
Second, the sensitivity to strain of the mode frequencies, particularly the
intermediate frequency A$_1$ mode, can provide an experimental probe of
the strain state of a BaTiO$_3$ layer.
Indeed, for the closely related material PbTiO$_3$, mode hardening has been observed in
films under compressive epitaxial stress \cite{sun}.

Although our focus here centers on the effects of strain,
macroscopic electric fields, interfacial effects, and finite
size effects must also play a fundamental role in determining the properties 
of coherent epitaxial thin films and superlattices. 
However, knowledge of strain effects has already proved to be invaluable 
for analyzing experimental and first-principles 
results \cite{us}, and for separating out the contributions of other intrinsic
as well as extrinsic effects. Therefore, a broader study of BaTiO$_3$,
which includes calculation of the 
dielectric tensor and related quantities, is currently in progress.
A similar study of other perovskites, such as PbTiO$_3$, PbZrO$_3$, SrTiO$_3$,
KNbO$_3$ and KTaO$_3$, would also be pertinent.

\section{ACKNOWLEDGMENTS}
This work was supported by NSF-NIRT and ONR N00014-00-1-0261.
We thank C. H. Ahn, M. H. Cohen, J. \'I\~niguez, X. Q. Pan, D. G. Schlom,
and D. Vanderbilt for valuable discussions.

%
\end{document}